\documentclass[prd,onecolumn,nofootinbib,showpacs,showkeys]{revtex4}
\newcommand\gothfamily{\usefont{U}{ygoth}{m}{n}}
\DeclareTextFontCommand{\textgoth}{\gothfamily}

\begin{document}

\title{GEOMETRIZATION OF ELECTROMAGNETISM IN TETRAD-SPIN-CONNECTION GRAVITY}
\author{{\bf Nikodem J. Pop\l awski}}

\affiliation{Department of Physics, Indiana University, Swain Hall West, 727 East Third Street, Bloomington, IN 47405, USA}
\email{nipoplaw@indiana.edu}

\noindent
{\em Modern Physics Letters A}\\
Vol. {\bf 24}, No. 6 (2009) 431--442\\
\copyright\,World Scientific Publishing Co.
\vspace{0.4in}

\begin{abstract}
The metric-affine Lagrangian of Ponomarev and Obukhov for the unified gravitational and electromagnetic field is linear in the Ricci scalar and quadratic in the tensor of homothetic curvature.
We apply to this Lagrangian the variational principle with the tetrad and spin connection as dynamical variables and show that, in this approach, the field equations are the Einstein-Maxwell equations if we relate the electromagnetic potential to the trace of the spin connection.
We also show that, as in the Ponomarev-Obukhov formulation, the generally covariant Dirac Lagrangian gives rise to the standard spinor source for the Einstein-Maxwell equations, while the spinor field obeys the nonlinear Heisenberg-Ivanenko equation with the electromagnetic coupling.
We generalize that formulation to spinors with arbitrary electric charges.
\end{abstract}

\pacs{04.20.Fy, 04.40.Nr, 04.50.Kd, 11.10.Ef}
\keywords{unified field theory; metric-affine gravity; homothetic curvature; spin connection; spinor connection.}

\maketitle

\section{Introduction}

In general relativity, the electromagnetic field and its sources are considered to be on the side of the matter tensor in the field equations, i.e. they act as sources of the gravitational field.
In unified field theory, the electromagnetic field obtains the same geometric status as the gravitational field~\cite{Goe}.
The geometry of general relativity is that of a four-dimensional Riemannian manifold, i.e. equipped with a symmetric metric-tensor field and an affine connection that is torsionless and metric compatible.
In order to combine gravitation and electromagnetism on the classical level within a geometrical theory we must modify some postulates of general relativity.
Weyl relaxed the postulate of metric compatibility of the affine connection, obtaining a unified theory of gravitation and electromagnetism, where gauge invariance of the electromagnetic potential is related to conformal invariance of the gravitational Lagrangian~\cite{Weyl1,Weyl2,Weyl3}.
Kaluza introduced a five-dimensional spacetime with one Killing vector and showed that the Lagrangian linear in the five-dimensional Ricci scalar yields the Einstein-Maxwell field equations and the Lorentz equation of motion~\cite{Kaluza}.
Relaxing the postulate of the symmetry of the affine connection~\cite{Car} and metric tensor resulted in the Einstein-Straus and Schr\"{o}dinger nonsymmetric unified field theories~\cite{Schrod1,Schrod2,Schrod3,Schrod4,Schrod5,Schrod6,Einst}.
While Kaluza's theory gave rise to later models of spacetime with extra dimensions, the Einstein-Schr\"{o}dinger and Weyl's theories turned out to be unphysical~\cite{Lord}.

In the {\em metric-affine} formulation of gravity, both the metric tensor and connection are independent variables (gravitational potentials) and the field equations are derived by varying the action with respect to these quantities~\cite{HD,MA1,MA2,MA3,MA4,MA5}.
A general affine connection has enough degrees of freedom to describe both the gravitational and electromagnetic field.
The gravitational field is represented in the Lagrangian by the symmetric part of the Ricci tensor and the classical electromagnetic field can be represented by the tensor of {\em homothetic curvature} (Weyl's segmental curvature tensor)~\cite{Scho}.
Unlike in the Einstein-Schr\"{o}dinger and Weyl's theories, where the electromagnetic field is associated with a generalized metric tensor, incorporating the electromagnetic potential into a generalized affine connection seems more natural: the connection generalizes an ordinary derivative of a vector into a coordinate-covariant derivative and the electromagnetic potential generalizes it into a $U(1)$-covariant derivative, so both objects have the same purpose: to preserve the correct transformation properties under certain symmetries.
The simplest metric-affine Lagrangian that depends on the Ricci scalar and the tensor of homothetic curvature (linear in the Ricci scalar and quadratic in the tensor of homothetic curvature), introduced by Ponomarev and Obukhov, generates the Einstein-Maxwell equations~\cite{PO}.
The analogous Lagrangian in the purely affine formulation was introduced by Ferraris and Kijowski~\cite{FK}.

In order to incorporate matter fields represented by spinors we must use a tetrad as a dynamical variable instead of the metric tensor.
Accordingly, the variation with respect to the affine connection can be replaced by the variation with respect to the spin connection (the Einstein-Cartan-Kibble-Sciama theory)~\cite{tetsp1,tetsp2,tetsp3,tetsp4,rev}.
In this paper, which follows Refs.~\cite{Niko1,Niko2}, we use the tetrad and spin connection as the gravitational potentials and show that the Lagrangian of Ponomarev and Obukhov~\cite{PO} generates in this formulation the Einstein-Maxwell equations with the electromagnetic potential represented by the trace of the spin connection.
We also show that the generally covariant Dirac Lagrangian produces in this tetrad-spin-connection formulation the standard spinor source for the Einstein-Maxwell equations, while the spinor field obeys the nonlinear Heisenberg-Ivanenko equation~\cite{HD,Heis1,Heis2,Heis3,grav1,grav2} with the electromagnetic coupling.
The model of Ponomarev and Obukhov~\cite{PO} describes spinors with the same (nonzero) electric charge, set by the constants in the Lagrangian.
We demonstrate that the nonuniqueness of how the spin connection enters the spinor connection allows to describe spinors with arbitrary electric charges, including zero.

\section{Variations of the Ricci scalar and homothetic curvature}

The curvature tensor with two Lorentz and two coordinate indices depends only on the {\em spin connection} $\omega^a_{\phantom{a}b\mu}$ and its first derivatives~\cite{Lord,Niko1}:
\begin{equation}
R^a_{\phantom{a}b\mu\nu}=\omega^a_{\phantom{a}b\nu,\mu}-\omega^a_{\phantom{a}b\mu,\nu}+\omega^a_{\phantom{a}c\mu}\omega^c_{\phantom{c}b\nu}-\omega^a_{\phantom{a}c\nu}\omega^c_{\phantom{c}b\mu}.
\label{curva1}
\end{equation}
The double contraction of the tensor~(\ref{curva1}) with the tetrad $e^\mu_a$ gives the Ricci scalar:
\begin{equation}
R=R^a_{\phantom{a}b\mu\nu}e^\mu_a e^{b\nu}.
\label{curva2}
\end{equation}
The tensor of homothetic curvature, $Q_{\mu\nu}=R^c_{\phantom{c}c\mu\nu}$, as a function of the spin connection is given by~\cite{Niko1}
\begin{equation}
Q_{\mu\nu}=\omega^c_{\phantom{c}c\nu,\mu}-\omega^c_{\phantom{c}c\mu,\nu}.
\label{second}
\end{equation}

The simplest metric-affine Lagrangian density that depends on the Ricci scalar and the tensor of homothetic curvature was introduced by Ponomarev and Obukhov~\cite{PO}:
\begin{equation}
\textgoth{L}=-\frac{\sqrt{-g}R}{2\kappa}+\frac{\sqrt{-g}\alpha^2}{4}Q_{\mu\nu}Q^{\mu\nu}+\textgoth{L}_m,
\label{Lagr}
\end{equation}
where $\textgoth{L}_m$ is the matter part of the Lagrangian density, $\alpha$ is a constant and $c=1$.
The gravitational part of $\textgoth{L}$ can be written as
\begin{equation}
\sqrt{-g}R=2\textgoth{E}^{\mu\nu}_{ab}(\omega^{ab}_{\phantom{ab}\nu,\mu}+\omega^a_{\phantom{a}c\mu}\omega^{cb}_{\phantom{cb}\nu}),
\label{Ric}
\end{equation}
where $\textgoth{E}^{\mu\nu}_{ab}=\textgoth{e}e^{[\mu}_a e^{\nu]}_b$ and $\textgoth{e}=\mbox{det}(e_\mu^a)$.
The variation of the Ricci scalar density with respect to the tetrad gives
\begin{equation}
\delta(\sqrt{-g}R)=(2R^a_\mu-Re^a_\mu)\textgoth{e}\delta e^\mu_a,
\label{var1}
\end{equation}
where $R^a_\mu=R^{[ab]}_{\phantom{[ab]}\mu\nu}e^\nu_b$ is the mixed (one Lorentz and one coordinate index) Ricci tensor.
The variation of the electromagnetic part of the Lagrangian density gives
\begin{equation}
\delta(\textgoth{e}Q_{\mu\nu}Q^{\mu\nu})=(4Q_{\mu\nu}Q_\rho^{\phantom{\rho}\nu}e^{a\rho}-Q_{\alpha\beta}Q^{\alpha\beta}e^a_\mu)\textgoth{e}\delta e^\mu_a.
\label{var2}
\end{equation}
Consequently, the stationarity of the action $S=\int d^4x\textgoth{L}$ under the variation of the tetrad yields the Einstein equations:
\begin{equation}
R^a_\mu-\frac{1}{2}Re^a_\mu=\kappa\alpha^2\bigl(Q_{\mu\nu}Q_\rho^{\phantom{\rho}\nu}e^{a\rho}-\frac{1}{4}Q_{\alpha\beta}Q^{\alpha\beta}e^a_\mu\bigr)+\frac{\kappa}{\textgoth{e}}\textgoth{T}^a_\mu,
\label{var3}
\end{equation}
where $\textgoth{T}^a_\mu$ is a dynamical energy-momentum tensor density in the tetrad formulation of gravity: $\delta\textgoth{L}_m=\textgoth{T}^a_\mu\delta e^\mu_a$~\cite{Lord,rev,Niko2}.

Varying the action with respect to the connection in metric-affine theories of gravity gives a relation between the connection and metric.
We obtain an analogous relation by varying the action with respect to the spin connection, related to the affine connection $\Gamma^{\,\,\rho}_{\mu\,\nu}$ by~\cite{tetsp1,tetsp2,tetsp3,tetsp4,rev,Niko1}
\begin{equation}
\omega^a_{\phantom{a}b\mu}=e^a_\nu e^\nu_{b,\mu}+e^a_\nu e^\rho_b\Gamma^{\,\,\nu}_{\rho\,\mu}.
\label{om1}
\end{equation}
The variation of the Ricci scalar density with respect to the spin connection gives
\begin{equation}
\delta(\sqrt{-g}R)=2\textgoth{E}^{\mu\nu}_{ab}\delta(\omega^{ab}_{\phantom{ab}\nu,\mu}+\omega^a_{\phantom{a}c\mu}\omega^{cb}_{\phantom{cb}\nu})=2(\textgoth{E}^{\mu\nu}_{ab,\nu}+\textgoth{E}^{\mu\nu}_{ac}\omega_{b\phantom{c}\nu}^{\phantom{b}c}-\textgoth{E}^{\mu\nu}_{cb}\omega^c_{\phantom{c}a\nu})\eta^{bc}\delta\omega^a_{\phantom{a}c\mu}.
\label{var4}
\end{equation}
The total (with respect to the coordinate, Lorentz and spinor indices) covariant derivative of the tensor density $\textgoth{E}^{\mu\nu}_{ab}$ vanishes due to vanishing of the total covariant derivative of the tetrad~\cite{Lord,Niko1}:
\begin{equation}
\textgoth{E}^{\mu\nu}_{ab|\nu}=\textgoth{E}^{\mu\nu}_{ab,\nu}-\omega^c_{\phantom{c}a\nu}\textgoth{E}^{\mu\nu}_{cb}-\omega^c_{\phantom{c}b\nu}\textgoth{E}^{\mu\nu}_{ac}+S^\mu_{\phantom{\mu}\rho\nu}\textgoth{E}^{\rho\nu}_{ab}+\Gamma^{\,\,\rho}_{\nu\,\rho}\textgoth{E}^{\mu\nu}_{ab}-\Gamma^{\,\,\rho}_{\rho\,\nu}\textgoth{E}^{\mu\nu}_{ab}=0,
\label{var5}
\end{equation}
where $S^\rho_{\phantom{\rho}\mu\nu}=\Gamma^{\,\,\,\,\rho}_{[\mu\,\nu]}$ is the Cartan torsion tensor.\footnote{
We omit total derivatives in the Lagrangian density since they do not contribute to the variation of the action.
}
From Eqs.~(\ref{var4}) and~(\ref{var5}), and the formula~\cite{Niko1}
\begin{equation}
\omega_{(ab)\mu}=-\frac{1}{2}N_{ab\mu},
\label{symet}
\end{equation}
where $N_{\mu\nu\rho}=g_{\mu\nu;\rho}$ is the nonmetricity tensor,\footnote{
Schouten~\cite{Scho} defines the nonmetricity tensor as $Q_{\rho\mu\nu}=-g_{\mu\nu;\rho}$.
}
we obtain
\begin{equation}
\delta(\sqrt{-g}R)=-2(S^\mu_{\phantom{\mu}\rho\nu}\textgoth{E}^{\rho\nu}_{ab}+2S_\nu\textgoth{E}^{\mu\nu}_{ab}+N_{b\phantom{c}\nu}^{\phantom{b}c}\textgoth{E}^{\mu\nu}_{ac})\eta^{bc}\delta\omega^a_{\phantom{a}c\mu},
\label{var6}
\end{equation}
where $S_\mu=S^\nu_{\phantom{\nu}\mu\nu}$ is the torsion vector~\cite{Niko2}.
The semicolon denotes the covariant derivative with respect to $\Gamma^{\,\,\rho}_{\mu\,\nu}$, acting on the coordinate indices.

The variation of the electromagnetic part of the Lagrangian density~(\ref{Lagr}) gives, after omitting a total derivative,\footnote{
The electromagnetic part of the Lagrangian density~(\ref{Lagr}) is quadratic with respect to the nonmetricity tensor since the tensor of homothetic curvature is linear in $N_{\mu\nu\rho}$~\cite{Niko1}.
}
\begin{equation}
\delta(\textgoth{e}Q_{\mu\nu}Q^{\mu\nu})=-4(\textgoth{e}Q^{\nu\mu})_{,\nu}\delta^c_a\delta\omega^a_{\phantom{a}c\mu}.
\label{var7}
\end{equation}
Consequently, the stationarity of the action under the variation of the spin connection yields the field equation:
\begin{equation}
S^\mu_{\phantom{\mu}ab}+S_b e^\mu_a-S_a e^\mu_b+\frac{1}{2}(N_{bc}^{\phantom{bc}c}e^\mu_a-N_{b\phantom{\mu}a}^{\phantom{b}\mu})-\frac{\kappa\alpha^2}{\textgoth{e}}(\textgoth{e}Q^{\nu\mu})_{,\nu}\eta_{ab}+\frac{\kappa}{2\textgoth{e}}\textgoth{S}_{ab}^{\phantom{ab}\mu}=0.
\label{var8}
\end{equation}
where $\textgoth{S}_a^{\phantom{a}b\mu}$ is the {\em hypermomentum} density in the spin-connection formulation of gravity~\cite{rev,Niko2}: $\delta\textgoth{L}_m=\frac{1}{2}\textgoth{S}_a^{\phantom{a}b\mu}\delta \omega^a_{\phantom{a}b\mu}$.

\section{The Maxwell equations}

Contracting the field equation~(\ref{var8}) with respect to the indices $(a,b)$ brings it into the Maxwell-like equation:
\begin{equation}
\textgoth{e}Q^{\nu\mu}_{\phantom{\nu\mu}:\nu}=(\textgoth{e}Q^{\nu\mu})_{,\nu}=\frac{1}{8\alpha^2}\textgoth{S}_c^{\phantom{c}c\mu},
\label{Max1}
\end{equation}
where the colon denotes the covariant derivative with respect to the Christoffel symbols $\{^{\,\,\rho}_{\mu\,\nu}\}$.
Since the tensor of homothetic curvature $Q_{\mu\nu}$ is also a curl of the vector $\omega^c_{\phantom{c}c\mu}$, we can associate it with the electromagnetic field tensor $F_{\mu\nu}$:
\begin{equation}
Q_{\mu\nu}=\frac{i}{\alpha}F_{\mu\nu}.
\label{Max2}
\end{equation}
Accordingly, the electromagnetic potential $A_\mu$ is related to the trace of the spin connection:
\begin{equation}
\omega^c_{\phantom{c}c\mu}=\frac{i}{\alpha}A_\mu.
\label{Max3}
\end{equation}
With this association the electromagnetic part of the Lagrangian density~(\ref{Lagr}) acquires the correct minus sign.
The reason for the imaginarity of $\omega^c_{\phantom{c}c\mu}$ will be explained in the next section.
The trace of the spin-connection hypermomentum density is the source for the homothetic field, so it is proportional to the electromagnetic current vector:
\begin{equation}
j^\mu=-\frac{i}{8\alpha\textgoth{e}}\textgoth{S}_c^{\phantom{c}c\mu}.
\label{Max4}
\end{equation}

Substituting Eq.~(\ref{Max1}) to~(\ref{var8}) gives
\begin{equation}
S^\mu_{\phantom{\mu}ab}+S_b e^\mu_a-S_a e^\mu_b+\frac{1}{2}(N_{bc}^{\phantom{bc}c}e^\mu_a-N_{b\phantom{\mu}a}^{\phantom{b}\mu})-\frac{\kappa}{8\textgoth{e}}\textgoth{S}_c^{\phantom{c}c\mu}\eta_{ab}+\frac{\kappa}{2\textgoth{e}}\textgoth{S}_{ab}^{\phantom{ab}\mu}=0.
\label{var9}
\end{equation}
This equation determines the torsion and nonmetricity tensors, thus the affine and spin connection, from the hypermomentum density and tetrad.
In the absence of spinor fields: $\textgoth{S}_{ab}^{\phantom{ab}\mu}=0$, Eq.~(\ref{var9}) yields
\begin{equation}
N_{bc}^{\phantom{bc}c}=-\frac{4}{3}S_b,\,\,\,\,N^c_{\phantom{c}ca}=-\frac{16}{3}S_a,
\label{N}
\end{equation}
in agreement with the relation $\Gamma^{\,\,\rho}_{\mu\,\nu}=\{^{\,\,\rho}_{\mu\,\nu}\}-\frac{2}{3}\delta^\rho_\mu S_\nu$, which is the solution of the field equation arising from the variation of the action (in the absence of matter) with respect to the affine connection $\Gamma^{\,\,\rho}_{\mu\,\nu}$~\cite{MA3,Niko0}.
Combining Eqs.~(\ref{symet}), (\ref{Max3}) and~(\ref{N}) gives
\begin{equation}
S_\mu=\frac{3i}{8\alpha}A_\mu,
\label{S}
\end{equation}
so the torsion vector is proportional to the electromagnetic potential~\cite{Ham} and imaginary~\cite{Kun}.
The case $S_\mu=0$ corresponds to general relativity, for which the spin connection $\omega^{ab}_{\phantom{ab}\mu}$ is antisymmetric in the indices $(a,b)$~\cite{Niko1}.

Metric-affine Lagrangians that depend explicitly on a general, unconstrained affine connection and the symmetric part of the Ricci tensor are subject to an unphysical constraint on the sources~\cite{San1,San2}.
Since the tensor of homothetic curvature is antisymmetric, the current vector density $\textgoth{e}j^\mu$ must be conserved: $(\textgoth{e}j^\mu)_{,\mu}=\textgoth{e}j^\mu_{\phantom{\mu}:\mu}=0$, which constrains how the spin connection can enter the metric-affine Lagrangian for matter $\textgoth{L}_m$: $\textgoth{S}_{c\phantom{c\mu},\mu}^{\phantom{c}c\mu}=0$.
If the matter Lagrangian does not depend on $Q_{\mu\nu}$, the conservation of the current $j^\mu$ becomes a stronger, algebraic constraint on the hypermomentum density: $\textgoth{S}_c^{\phantom{c}c\mu}=0$.
The dependence of metric-affine Lagrangians on the tensor of homothetic curvature replaces this constraint with the field equation for $Q_{\mu\nu}$ that we associate with the Maxwell equation for the electromagnetic field.

The Ricci scalar is invariant under a projective transformation:
\begin{equation}
\omega^a_{\phantom{a}b\mu}\rightarrow\omega^a_{\phantom{a}b\mu}+\delta^a_b V_\mu,
\label{proj1}
\end{equation}
where $V_\mu$ is an arbitrary vector.
Under the same transformation, the tensor of homothetic curvature changes according to $Q_{\mu\nu}\rightarrow Q_{\mu\nu}+4(V_{\nu,\mu}-V_{\mu,\nu})$.
Consequently, the total action changes by $\delta S=\int d^4x(\frac{1}{2}\textgoth{S}_{ab}^{\phantom{ab}\mu}\delta \omega^{ab}_{\phantom{ab}\mu}+\frac{\textgoth{e}\alpha^2}{2}Q^{\mu\nu}\delta Q_{\mu\nu})=\int d^4x(\frac{1}{2}\textgoth{S}_{ab}^{\phantom{ab}\mu}\eta^{ab}V_\mu+4\textgoth{e}\alpha^2 Q^{\mu\nu}V_{\nu,\mu})$.
This expression is identically zero due to the field equation~(\ref{Max1}) so the action is projectively invariant.
Therefore we can interpret the electromagnetic field in metric-affine gravity as the field whose role is to preserve the projective invariance of metric-affine Lagrangians that depend explicitly on the affine connection without constraining the connection~\cite{Niko0}.

Although the tensor of homothetic curvature is not invariant under general projective transformations~(\ref{proj1}), it is invariant under special projective transformations:
\begin{equation}
\omega^a_{\phantom{a}b\mu}\rightarrow\omega^a_{\phantom{a}b\mu}+\delta^a_b\lambda_{,\mu},
\label{proj2}
\end{equation}
that correspond to $\lambda$-transformations of the affine connection~\cite{Einst}.
Because of Eq.~(\ref{Max3}) the transformation~(\ref{proj2}) is a geometric representation of the gauge transformation of the electromagnetic potential:
\begin{equation}
A_\mu \rightarrow A_\mu+\phi_{,\mu},
\label{gauge1}
\end{equation}
where
\begin{equation}
\phi=-4i\alpha\lambda.
\label{gauge2}
\end{equation}
In the presence of the gravitational field we ``correct'' the derivative by introducing the affine connection, while in the presence of the electromagnetic field we introduce the electromagnetic potential.
Therefore it seems natural to assume that the electromagnetic potential is related to the connection~\cite{PO} rather than to the metric as in earlier unified field theories~\cite{Weyl1,Weyl2,Weyl3,Kaluza,Schrod1,Schrod2,Schrod3,Schrod4,Schrod5,Schrod6,Weyl4}.

\section{The Heisenberg-Ivanenko equation}

The Dirac Lagrangian density for a spinor field $\psi$ (representing matter) with mass $m$ in the presence of the gravitational field is given by
\begin{equation}
\textgoth{L}_m=\frac{i\textgoth{e}}{2}(\bar{\psi}\gamma^\mu \psi_{|\mu}-\bar{\psi}_{|\mu}\gamma^\mu \psi)-\textgoth{e}m\bar{\psi}\psi=\frac{i\textgoth{e}}{2}(\bar{\psi}\gamma^\mu \psi_{,\mu}-\bar{\psi}_{,\mu}\gamma^\mu \psi)-\frac{i\textgoth{e}}{2}\bar{\psi}\{\gamma^\mu,\Gamma_\mu\}\psi-\textgoth{e}m\bar{\psi}\psi,
\label{Dir1}
\end{equation}
where $\bar{\psi}=\psi^\dagger\gamma^0$ is the adjoint spinor corresponding to $\psi$, $\Gamma_\mu$ is the {\em spinor connection}~\cite{Lord,rev,Niko1}, $\{\}$ denotes anticommutation, and $\hbar=1$.
In the presence of nonmetricity, the spinor connection is given by the Fock-Ivanenko coefficients with the antisymmetric part of the spin connection~\cite{Niko1,sp}:\footnote{
This form of the spinor connection results from
\begin{equation}
\gamma^a_{\phantom{a}|\mu}=\omega^a_{\phantom{a}b\mu}\gamma^b-[\Gamma_\mu,\gamma^a]=-\frac{1}{2}N^a_{\phantom{a}b\mu}\gamma^b.
\label{FI}
\end{equation}
}
\begin{equation}
\Gamma_\mu=-\frac{1}{4}\omega_{[ab]\mu}\gamma^a \gamma^b.
\label{Dir2}
\end{equation}
The electromagnetic field, like the Weyl conformal vector~\cite{Lord}, seems not to couple to the Dirac spinor because only the antisymmetric part of the spin connection appears in the Lagrangian density~(\ref{Dir1}).
The corresponding spin-connection hypermomentum density $\textgoth{S}_{ab}^{\phantom{ab}\mu}$ is antisymmetric in the indices $(a,b)$ and the current $j^\mu$ vanishes due to Eq.~(\ref{Max4}).

However, Eq.~(\ref{Dir2}) is not a unique solution for the spinor connection in Eq.~(\ref{FI}); we can add to the Fock-Ivanenko coefficients an arbitrary vector multiple of the unit matrix, $V_\mu$~\cite{Niko1}:
\begin{equation}
\Gamma_\mu=-\frac{1}{4}\omega_{[ab]\mu}\gamma^a \gamma^b+V_\mu.
\label{Dir3}
\end{equation}
Though only the antisymmetric part of the spin connection appears explicitly in Eq.~(\ref{Dir3}), we can insert the symmetric part $\omega_{(ab)\mu}$ into Eq.~(\ref{Dir3}) by choosing
\begin{equation}
V_\mu=-\frac{q}{4}\omega_{(ab)\mu}\gamma^a \gamma^b=-\frac{q}{4}\omega^c_{\phantom{c}c\mu},
\label{Dir4}
\end{equation}
where $q$ is a number related, as we show below, to the electric charge of the spinor.
We can rewrite Eq.~(\ref{Dir3}) as
\begin{equation}
\Gamma_\mu^{(q)}=-\frac{1}{4}\omega^{(q)}_{ab\mu}\gamma^a \gamma^b,
\label{Dir5}
\end{equation}
where
\begin{equation}
\omega^{(q)}_{ab\mu}=\omega_{[ab]\mu}+q\omega_{(ab)\mu}
\label{Dir6}
\end{equation}
is the modified spin connection.\footnote{
The nonuniqueness of the spinor connection up to a vector $V_\mu$ allows to introduce gauge fields interacting with spinors~\cite{Niko1}.
}
Since the modified spinor connection $\Gamma_\mu^{(q)}$ corresponding to the addition of the vector~(\ref{Dir4}) is related to the Fock-Ivanenko spinor connection $\Gamma_\mu$ by
\begin{equation}
\Gamma_\mu^{(q)}=\Gamma_\mu-\frac{q}{4}\omega^c_{\phantom{c}c\mu},
\label{Dir7}
\end{equation}
we must add to the right-hand side of Eq.~(\ref{Dir1}) the term
\begin{equation}
\textgoth{L}_{(q)}=\frac{iq\textgoth{e}}{4}\omega^c_{\phantom{c}c\mu}\bar{\psi}\gamma^\mu\psi.
\label{Dir8}
\end{equation}

The part of the matter Lagrangian density~(\ref{Dir1}) that contains the spin connection is given by
\begin{equation}
\textgoth{L}_\omega=-\frac{i\textgoth{e}}{2}\bar{\psi}\{\gamma^\mu,\Gamma_\mu^{(q)}\}\psi=\frac{i\textgoth{e}}{4}\bar{\psi}\gamma^{[a}\gamma^b\gamma^{\mu]}\psi\omega_{ab\mu}+\textgoth{L}_{(q)},
\label{Dir9}
\end{equation}
where we used the identity $\{\gamma^a,\gamma^{[b}\gamma^{c]}\}=2\gamma^{[a}\gamma^b\gamma^{c]}$.
Consequently, the hypermomentum density is~\cite{grav1,grav2}
\begin{equation}
\textgoth{S}^{ab\mu}=\frac{i\textgoth{e}}{2}\bar{\psi}\gamma^{[a}\gamma^b\gamma^{\mu]}\psi+\frac{iq\textgoth{e}}{2}\bar{\psi}\gamma^\mu\psi\eta^{ab},
\label{Dir10}
\end{equation}
and Eq.~(\ref{var9}) becomes
\begin{equation}
S^\mu_{\phantom{\mu}ab}+S_b e^\mu_a-S_a e^\mu_b+\frac{1}{2}(N_{bc}^{\phantom{bc}c}e^\mu_a-N_{b\phantom{\mu}a}^{\phantom{b}\mu})+\frac{i\kappa}{4}e^{\mu c}\bar{\psi}\gamma_{[a}\gamma_b\gamma_{c]}\psi=0.
\label{Dir11}
\end{equation}
Using the relations~(\ref{N}), which remain valid in the presence of the Dirac spinors, we obtain
\begin{equation}
S^\mu_{\phantom{\mu}ab}-\frac{1}{4}N_{bc}^{\phantom{bc}c}e^\mu_a
+\frac{3}{4}N_{ac}^{\phantom{ac}c}e^\mu_b-\frac{1}{2}N_{b\phantom{\mu}a}^{\phantom{b}\mu}+\frac{i\kappa}{4}e^{\mu c}\bar{\psi}\gamma_{[a}\gamma_b\gamma_{c]}\psi=0.
\label{Dir12}
\end{equation}

Relating the constant $\alpha$ to charge of the electron $e$ by $\alpha=\frac{1}{4e}$ and using Eq.~(\ref{Max3}) bring Eq.~(\ref{Max4}) to the form: $j^\mu=qe\bar{\psi}\gamma^\mu\psi$, and Eq.~(\ref{Dir8}) to $\textgoth{L}_{(q)}=-\textgoth{e}A_\mu j^\mu$.\footnote{
The constant $\alpha$ has the dimension of the electric charge, which follows from how it appears in the Lagrangian density~(\ref{Lagr}).
}
Therefore $q$ represents the {\em electric charge} (in units of the charge of the electron) associated with the spinor $\psi$.
Equation~(\ref{Dir7}) takes the form: $\Gamma_\mu^{(q)}=\Gamma_\mu-iqeA_\mu$, i.e. the covariant derivative arising from the electromagnetic field represented by the tensor of homothetic curvature coincides with the covariant derivative of the $U(1)$ gauge symmetry, explaining why in Eq.~(\ref{Max3}) we associated the spin connection with the electromagnetic potential multiplied by $i$.

The invariance of Eq.~(\ref{FI}) under the addition of a vector multiple $V_\mu$ of the unit matrix to the spinor connection allows to introduce the interaction between spinors and the vector field $V_\mu$~\cite{Niko1}.
Rewriting $V_\mu$ in terms of the symmetric part of the spin connection, as in Eq.~(\ref{Dir4}), gives the proportionality constant $q$ the interpretation of the coupling between the spinor field and $V_\mu$, i.e. the electric charge of the spinor.
The nonuniqueness of the spinor connection up to a vector is thus related to the arbitrariness of the electric charge of a (classical) spinor field.
The quantization of the electric charge may result from the quantization of the modified spin connection~(\ref{Dir6}).
In the model of Ponomarev and Obukhov~\cite{PO}, the spinor connection is set to $\Gamma_\mu=-\frac{1}{4}\omega_{ab\mu}\gamma^a \gamma^b$, which differs from Eq.~(\ref{Dir2}) by a vector multiple of the unit matrix and corresponds to spinors with $q=1$.
Here we show that the metric-affine formulation of gravity and electromagnetism with the electromagnetic field represented by the tensor of homothetic curvature allows to describe spinors with arbitrary electric charges, including electrically neutral spinors.

Varying the action with respect to the spinor fields gives the field equations for spinors.
The variation of the Dirac Lagrangian density~(\ref{Dir1}) with respect to $\psi$ gives\footnote{
We again omit total derivatives in the Lagrangian density.
}
\begin{equation}
\delta\textgoth{L}_m=-\Bigl(\frac{i}{2}\bigl((\textgoth{e}\bar{\psi}\gamma^\mu)_{,\mu}+\textgoth{e}\bar{\psi}\gamma^\mu\Gamma_\mu+\textgoth{e}\bar{\psi}_{|\mu}\gamma^\mu\bigr)+\textgoth{e}m\bar{\psi}\Bigr)\delta\psi,
\label{HI1}
\end{equation}
while the variation with respect to $\bar{\psi}$ gives the equivalent equation:
\begin{equation}
\delta\textgoth{L}_m=\delta\bar{\psi}\Bigl(\frac{i}{2}\bigl((\textgoth{e}\gamma^\mu\psi)_{,\mu}-\textgoth{e}\Gamma_\mu\gamma^\mu\psi+\textgoth{e}\gamma^\mu\psi_{|\mu}\bigr)-\textgoth{e}m\psi\Bigr).
\label{HI2}
\end{equation}
Equation~(\ref{HI2}) yields the field equation for $\psi$:
\begin{equation}
2\textgoth{e}\gamma^\mu\psi_{|\mu}+\textgoth{e}\gamma^\mu\Gamma_\mu\psi-\textgoth{e}\Gamma_\mu\gamma^\mu\psi+(\textgoth{e}\gamma^\mu)_{;\mu}\psi-2\textgoth{e}S_\mu\gamma^\mu\psi+2i\textgoth{e}m\psi=0.
\label{HI3}
\end{equation}
Using the formulae: $\gamma^\mu_{\phantom{\mu}|\nu}=\gamma^\mu_{\phantom{\mu};\nu}-[\Gamma_\nu,\gamma^\mu]$, $\gamma^\mu_{\phantom{\mu}|\nu}=-\frac{1}{2}N^\mu_{\phantom{\mu}\rho\nu}\gamma^\rho$ and $\textgoth{e}_{;\mu}=\frac{\textgoth{e}}{2}N^\nu_{\phantom{\nu}\nu\mu}$~\cite{Niko1}, where $[\,]$ denotes commutation, and replacing $\Gamma_\mu$ by $\Gamma_\mu^{(q)}$, we can write Eq.~(\ref{HI3}) as
\begin{equation}
\gamma^\mu\psi_{,\mu}-\gamma^\mu\Gamma_\mu^{(q)}\psi-S_\mu\gamma^\mu\psi+\frac{1}{2}N^\nu_{\phantom{\nu}[\nu\mu]}\gamma^\mu\psi+im\psi=0.
\label{HI4}
\end{equation}
Combining Eqs.~(\ref{N}) and~(\ref{HI4}) gives
\begin{equation}
\gamma^\mu(\psi_{,\mu}+iqeA_\mu\psi+\frac{1}{4}\omega_{[ab]\mu}\gamma^a\gamma^b\psi)+\frac{3}{2}N_{\mu c}^{\phantom{\mu c}c}\gamma^\mu\psi+im\psi=0.
\label{HI5}
\end{equation}

The spin connection $\omega_{ab\mu}$ is related to the affine connection by Eq.~(\ref{om1}), and the affine connection is related to the torsion and nonmetricity tensors by~\cite{Scho}
\begin{equation}
\Gamma^{\,\,\rho}_{\mu\,\nu}=\{^{\,\,\rho}_{\mu\,\nu}\}+S^\rho_{\phantom{\rho}\mu\nu}+S_{\mu\nu}^{\phantom{\mu\nu}\rho}+S_{\nu\mu}^{\phantom{\nu\mu}\rho}+\frac{1}{2}(N_{\mu\nu}^{\phantom{\mu\nu}\rho}-N^\rho_{\phantom{\rho}\mu\nu}-N^\rho_{\phantom{\rho}\nu\mu}).
\label{om2}
\end{equation}
Therefore the spin connection $\omega_{ab\mu}$ can be written as the sum of the antisymmetric spin connection $\omega_{ab\mu}^{\{\}}$ corresponding to the symmetric and metric-compatible Levi-Civita affine connection $\{^{\,\,\rho}_{\mu\,\nu}\}$ (we denote the corresponding spinor connection by $\Gamma_\mu^{\{\}}$):
\begin{equation}
\omega_{ab\mu}^{\{\}}=e_{a\nu}e^\nu_{b,\mu}+e_{a\nu}e^\rho_b\{^{\,\,\nu}_{\rho\,\mu}\},
\label{om3}
\end{equation}
and the non-Riemannian part $\omega_{ab\mu}^{(n)}$ containing the torsion and nonmetricity tensors:
\begin{equation}
\omega_{ab\mu}^{(n)}=e_{a\nu}e^\rho_b\Bigl(S^\nu_{\phantom{\nu}\rho\mu}+S_{\rho\mu}^{\phantom{\rho\mu}\nu}+S_{\mu\rho}^{\phantom{\mu\rho}\nu}+\frac{1}{2}(N_{\rho\mu}^{\phantom{\rho\mu}\nu}-N^\nu_{\phantom{\nu}\rho\mu}-N^\nu_{\phantom{\nu}\mu\rho})\Bigr).
\label{om4}
\end{equation}

Substituting Eq.~(\ref{Dir12}) to~(\ref{om4}) and antisymmetrizing with respect to the indices $(a,b)$ gives
\begin{equation}
\omega_{[ab]\mu}^{(n)}=N_{ac}^{\phantom{ac}c}e_{b\mu}-N_{bc}^{\phantom{bc}c}e_{a\mu}-\frac{i\kappa}{4}\bar{\psi}\gamma_{[a}\gamma_b\gamma_{\mu]}\psi,
\label{om5}
\end{equation}
which, using the identity $\gamma^{[a}\gamma^b\gamma^{c]}=i\epsilon^{abcd}\gamma_d\gamma^5$, reads
\begin{equation}
\frac{1}{4}\gamma^\mu\gamma^a\gamma^b\omega_{[ab]\mu}^{(n)}=-\frac{3}{2}\gamma^\mu N_{\mu c}^{\phantom{\mu c}c}-\frac{3i\kappa}{8}(\bar{\psi}\gamma_\mu\gamma^5\psi)\gamma^\mu\gamma^5.
\label{om6}
\end{equation}
Combining Eqs.~(\ref{HI5}) and~(\ref{om6}) gives the Heisenberg-Ivanenko equation with the electromagnetic coupling~\cite{grav1,grav2}:
\begin{equation}
\gamma^\mu(\psi_{,\mu}-\Gamma_\mu^{\{\}}\psi+iqeA_\mu\psi)-\frac{3i\kappa}{8}(\bar{\psi}\gamma_\mu\gamma^5\psi)\gamma^\mu\gamma^5\psi+im\psi=0.
\label{HI6}
\end{equation}
Similarly, Eq.~(\ref{HI1}) yields
\begin{equation}
(\bar{\psi}_{,\mu}+\bar{\psi}\Gamma_\mu^{\{\}}-iqeA_\mu\bar{\psi})\gamma^\mu+\frac{3i\kappa}{8}\bar{\psi}\gamma^\mu\gamma^5(\bar{\psi}\gamma_\mu\gamma^5\psi)-im\bar{\psi}=0.
\label{HI7}
\end{equation}
The first terms on the left-hand sides of Eqs.~(\ref{HI6}) and~(\ref{HI7}) correspond to the general-relativistic interaction of the spinors $\psi$ and $\bar{\psi}$, respectively, with the electromagnetic potential $A_\mu$ (represented geometrically by the trace of the spin connection).
The second terms, nonlinear in $\psi$ or $\bar{\psi}$, respectively, describe the Heisenberg-Ivanenko spinor self-interaction that introduces deviations from the Dirac equation at energies on the order of the Planck energy, and may be related to the weak interaction of particles~\cite{HD,non1,non2,non3,non4}.
The homothetic curvature associated with the electromagnetic field does not introduce, besides generating the coupling between the electromagnetic potential and spinor, any changes to the Heisenberg-Ivanenko equation.
The same result is found if we vary the action with respect to the affine connection instead of the spin connection~\cite{PO}.

\section{Concluding remarks}

For a linear connection, not restricted to be metric compatible and symmetric, there are five possible modifications of the Maxwell equations: $g^{\nu\rho}F_{\nu\mu;\rho}=j_\mu$, $F^\nu_{\phantom{\nu}\mu;\nu}=j_\mu$, $F^{\nu\mu}_{\phantom{\nu\mu};\nu}=j^\mu$, $g^{\nu\rho}F_{\nu\phantom{\mu};\rho}^{\phantom{\nu}\mu}=j^\mu$, and the metrically modified $g^{\nu\rho}F_{\nu\mu:\rho}=j_\mu$~\cite{Coley}.\footnote{
The semicolon and colon denote the covariant derivatives with respect to $\Gamma^{\,\,\rho}_{\mu\,\nu}$ and $\{^{\,\,\rho}_{\mu\,\nu}\}$, respectively.
}
The experimentally confirmed conservation laws of electric charge and magnetic flux indicate that the last possibility, which also results from the differential-form and metric-free formulations of electrodynamics, is physical~\cite{Van1,Van2}.
Regarding the tensor of homothetic curvature $Q_{\mu\nu}$ as the geometrical quantity representing the electromagnetic field tensor in the metric-affine gravity, together with the simplest form of the Lagrangian that contains $Q_{\mu\nu}$~\cite{PO}, automatically leads to the metrically modified Maxwell equations.
In this paper we obtained the same result using the tetrad-spin-connection formulation (variation with respect to $\omega^a_{\phantom{a}b\mu}$ instead of $\Gamma^{\,\,\rho}_{\mu\,\nu}$).
We applied this formulation to the generally covariant Dirac Lagrangian and showed that the nonuniqueness of how the spin connection enters the spinor connection allows one to describe (nonlinear)~\cite{Ivan} spinors with arbitrary electric charges, generalizing the results of Ref.~\cite{PO}.

In the presence of the gravitational field we generalize an ordinary derivative into a coordinate-covariant derivative by introducing the affine connection, while in the presence of the electromagnetic field we generalize it into a $U(1)$-covariant derivative by introducing the electromagnetic potential.
In order to reproduce correctly the $U(1)$-covariant derivative, the tensor of homothetic curvature and the torsion vector must be purely imaginary because the electromagnetic-field observables are real.
Accordingly, the antisymmetric part of the spin connection is real, while its symmetric part is imaginary (Ref.~\cite{PO} does not discuss this point).
Relating electromagnetism to the affine connection~\cite{PO} or spin connection (as in this paper) seems more natural than associating it with the metric, as in earlier unified theories~\cite{Weyl1,Weyl2,Weyl3,Kaluza,Schrod1,Schrod2,Schrod3,Schrod4,Schrod5,Schrod6}.
This relation may also suggest the correct way of quantizing the gravitational field, since we we already have a highly successful quantum theory of the electromagnetic field (QED).
A successful theory unifying gravitational and electromagnetic interactions on the classical level should be regarded as the classical limit of the quantum theory of all interactions, giving insights on how to construct such a theory, possibly with geometrization of spinor fields.
Therefore classical unified field theory is still a topic worthy of investigation.


\begin{thebibliography}{}
\bibitem{Goe} H. F. M. Goenner, {\em Liv. Rev. Relativ.} {\bf 7}, 2 (2004).
\bibitem{Weyl1} H. Weyl, {\it Sitzungsber. Preuss. Akad. Wiss.} (Berlin), 465 (1918).
\bibitem{Weyl2} H. Weyl, {\it Ann. Phys.} (Leipzig) {\bf 59}, 101 (1919).
\bibitem{Weyl3} H. Weyl, {\em Space, Time, Matter} (Methuen, 1922).
\bibitem{Kaluza} T. Kaluza, {\it Sitzungsber. Preuss. Akad. Wiss.} (Berlin), 966 (1921).
\bibitem{Car} \'{E}. Cartan, {\it Compt. Rend. Acad. Sci.} (Paris) {\bf 174}, 593 (1922).
\bibitem{Schrod1} A. Einstein and E. G. Straus, {\em Ann. Math.} {\bf 47}, 731 (1946).
\bibitem{Schrod2} E. Schr\"{o}dinger, {\em Proc. R. Ir. Acad. A}, {\bf 51} 147 (1947).
\bibitem{Schrod3} E. Schr\"{o}dinger, {\em Proc. R. Ir. Acad. A} {\bf 51}, 163 (1947).
\bibitem{Schrod4} A. Einstein, {\it Rev. Mod. Phys.} {\bf 20}, 35 (1948).
\bibitem{Schrod5} E. Schr\"{o}dinger, {\em Space-Time Structure} (Cambridge Univ. Press, 1950).
\bibitem{Schrod6} M. A. Tonnelat, {\em Einstein's Unified Field Theory} (Gordon and Breach, 1966).
\bibitem{Einst} A. Einstein, {\it The Meaning of Relativity} (Princeton Univ. Press, 1956).
\bibitem{Lord} E. A. Lord, {\em Tensors, Relativity and Cosmology} (McGraw-Hill, 1976).
\bibitem{HD} F. W. Hehl and B. K. Datta, {\em J. Math. Phys.} {\bf 12}, 1334 (1971).
\bibitem{MA1} A. Einstein, {\it Sitzungsber. Preuss. Akad. Wiss.} (Berlin), 414 (1925).
\bibitem{MA2} L. L. Smalley, {\em Phys. Lett. A} {\bf 61}, 436 (1977).
\bibitem{MA3} F. W. Hehl and G. D. Kerlick, {\em Gen. Relativ. Gravit.} {\bf 9}, 691 (1978).
\bibitem{MA4} F. W. Hehl, E. A. Lord and L. L. Smalley, {\em Gen. Relativ. Gravit.} {\bf 13}, 1037 (1981).
\bibitem{MA5} F. W. Hehl, J. D. McCrea, E. W. Mielke and Y. Ne'eman, {\em Phys. Rep.} {\bf 258}, 1 (1995).
\bibitem{Scho} J. A. Schouten, {\em Ricci-Calculus} (Springer-Verlag, 1954).
\bibitem{PO} V. N. Ponomarev and Yu. N. Obukhov, {\em Gen. Relativ. Gravit.} {\bf 14}, 309 (1982).
\bibitem{FK} M. Ferraris and J. Kijowski, {\it Gen. Relativ. Gravit.} {\bf 14}, 37 (1982).
\bibitem{tetsp1} R. Utiyama, {\em Phys. Rev.} {\bf 101}, 1597 (1956).
\bibitem{tetsp2} T. W. B. Kibble, {\em J. Math. Phys.} {\bf 2}, 212 (1961).
\bibitem{tetsp3} D. W. Sciama, in {\em Recent Developments in General Relativity}, p. 415 (Pergamon, 1962).
\bibitem{tetsp4} D. W. Sciama, {\em Rev. Mod. Phys.} {\bf 36}, 463 (1964).
\bibitem{rev} F. W. Hehl, P. von der Heyde, G. D. Kerlick and J. M. Nester, {\em Rev. Mod. Phys.} {\bf 48}, 393 (1976).
\bibitem{Niko1} N. J. Pop\l awski, arXiv: 0710.3982 [gr-qc].
\bibitem{Niko2} N. J. Pop\l awski, arXiv: 0711.2341 [gr-qc].
\bibitem{Heis1} W. Heisenberg, {\em Rev. Mod. Phys.} {\bf 29}, 269 (1957).
\bibitem{Heis2} W. Heisenberg, {\it Introduction to the Unified Field Theory of Elementary Particles} (Interscience, 1966).
\bibitem{Heis3} D. Ivanenko and G. Sardanashvily, {\em Phys. Rep.} {\bf 94}, 1 (1983).
\bibitem{grav1} V. de Sabbata and M. Gasperini, {\em Introduction to Gravitation} (World Scientific, 1985).
\bibitem{grav2} V. de Sabbata and C. Sivaram, {\em Spin and Torsion in Gravitation} (World Scientific, 1994).
\bibitem{Niko0} N. J. Pop\l awski, {\em Int. J. Mod. Phys. A} {\bf 23}, 567 (2008).
\bibitem{Ham} R. T. Hammond, {\em Class. Quantum Grav.} {\bf 6}, L195 (1989).
\bibitem{Kun} G. Kunstatter, {\em Gen. Relativ. Gravit.} {\bf 12}, 373 (1980).
\bibitem{San1} V. D. Sandberg, {\em Phys. Rev. D} {\bf 12}, 3013 (1975).
\bibitem{San2} A. Papapetrou and J. Stachel, {\em Gen. Relativ. Gravit.} {\bf 9}, 1075 (1978).
\bibitem{Weyl4} J. Pullin and O. Bressan, {\em Gen. Relativ. Gravit.} {\bf 19}, 601 (1987).
\bibitem{sp} R. F. Bilyalov, {\em Russ. Phys. J.} {\bf 41}, 1134 (1998).
\bibitem{non1} N. N. Zao, {\em Russ. Phys. J.} {\bf 17}, 1746 (1974).
\bibitem{non2} B. V. Danilyuk, {\em Russ. Phys. J.} {\bf 20}, 1047 (1977).
\bibitem{non3} G. I. Shipov, {\em Russ. Phys. J.} {\bf 20}, 378 (1977).
\bibitem{non4} M. Novello, {\em Europhys. Lett.} {\bf 80}, 41001 (2007).
\bibitem{Coley} A. A. Coley, {\em Phys. Rev. D} {\bf 27}, 728 (1983).
\bibitem{Van1} M. A. Vandyck, {\em J. Phys. A} {\bf 29}, 2245 (1996).
\bibitem{Van2} R. A. Puntigam, C. L\"{a}mmerzahl and F. W. Hehl, {\em Class. Quantum Grav.} {\bf 14}, 1347 (1997).
\bibitem{Ivan} D. Ivanenko, {\em Russ. Phys. J.} {\bf 8}, 1 (1965).
\end{thebibliography}
\end{document}